\documentclass[10pt, conference]{IEEEtran}
\IEEEoverridecommandlockouts

\def\BibTeX{{\rm B\kern-.05em{\sc i\kern-.025em b}\kern-.08em
    T\kern-.1667em\lower.7ex\hbox{E}\kern-.125emX}}

\usepackage{graphicx} 
\usepackage{xcolor}
\usepackage{url}
\usepackage{booktabs}
\usepackage{multirow}
\usepackage{tcolorbox}
\usepackage{enumitem}
\usepackage{threeparttable}
\usepackage{soul}
\usepackage{seqsplit} 
\usepackage{amsmath}
\usepackage{amssymb}
\usepackage{tabularx}
\usepackage{ragged2e}
\usepackage[numbers,sort&compress]{natbib}

\usepackage{hyperref}

  \usepackage{pifont}

\usepackage{framed}

  {\endMakeFramed}

\begin{document}
\title{Ethics by Configuration: AI Ethics Principles in Coding Agent Instructions}


\author{
    \IEEEauthorblockN{Haoyu Gao}
    \IEEEauthorblockA{
    The University of Melbourne\\
    Victoria, Australia\\
    haoyug1@student.unimelb.edu.au}
    \and
    \IEEEauthorblockN{Peerachai Banyongrakkul}
    \IEEEauthorblockA{
    The University of Melbourne\\
    Victoria Australia\\
    peerachai.banyongrakkul@student.unimelb.edu.au}
    \and
    \IEEEauthorblockN{Marc Cheong}
    \IEEEauthorblockA{
    The University of Melbourne\\
    Victoria, Australia\\
    marc.cheong@unimelb.edu.au}
    \and
 \IEEEauthorblockN{Christoph Treude}
    \IEEEauthorblockA{
    Singapore Management University\\
    Singapore\\
    ctreude@smu.edu.sg}
    \and
    \IEEEauthorblockN{Sebastian Baltes}
    \IEEEauthorblockA{
     Heidelberg University\\
     Heidelberg, Germany\\
    sebastian.baltes@uni-heidelberg.de}
    \and
    \IEEEauthorblockN{Mansooreh Zahedi}
    \IEEEauthorblockA{
    The University of Melbourne\\
    Victoria, Australia\\
    mansooreh.zahedi@unimelb.edu.au}
  }

\maketitle

\begin{abstract}
As AI coding agents become integral to software development, projects increasingly utilise configuration files, such as context files and skills, to govern agent behaviour. While AI ethics principles are widely used in research and policy,  little is known about whether and how they are reflected when developers configure coding agents. We studied ethics principles in 40,079 configuration files across 8,893 GitHub repositories. Using a validated large language model (LLM) classification, we measured the prevalence of ethics-related statements, and qualitatively analysed 702 stratified samples across 679 files to uncover the ethical considerations reflected in these files and the parties they concern. Ethics-related statements appeared in 39.4\% of files and 61.5\% of repositories; concentrated on accountability, transparency, privacy protection, and reliability; and occurred more often in task-specific files such as skills than in project-wide context files. Our qualitative analysis identified 30 categories of ethical considerations, which take the form of concrete rules for the agent. Accountability appeared mainly as requirements for human approval, transparency as instructions to explain and honestly report the agent's work, and fairness as accessibility and localisation. These considerations reveal a difference in focus. Whereas public ethics frameworks aim to protect society and end users, configuration files mainly concern developers, maintainers, and community members in the development workflow. Ethics-related content appeared mainly in prose-based files, and whether agents follow these instructions remains untested. We discuss the implications for practitioners, researchers, and policymakers.
\end{abstract}

\section{Introduction}

With the rise of AI coding agents, developers increasingly use them across software development activities, including requirements analysis, code writing, test generation, and documentation maintenance~\cite{liu2024large, li2025rise}, with autonomy ranging from close human supervision to independent operation~\cite{feng2025levels}. Their outputs, from intermediate responses to final deliverables, can impact stakeholders across the software development life cycle (SDLC). For example, hallucinated or overly long responses can frustrate the developer working with the agent~\cite{martinez2025emotional}, while end users may benefit from added accessibility features~\cite{mowar2025codea11y}. Meanwhile, agent configuration files, such as \texttt{AGENTS.md}~\cite{chatlatanagulchai2025agent} and skills~\cite{gao2026registry}, provide additional context and guidance for the agent to achieve specific tasks. Against the backdrop of Agentic AI governance, ethics, and risk management -- to ``effectively deploy and de-risk the powerful capabilities of the technology'' \cite{Governance-Institute-of-Australia2026-bv} --  it is increasingly important to understand how ethical considerations are reflected in these configurations, which guide agent behaviour in software development.

AI ethics principles, such as fairness and transparency, are widely discussed in research and policy~\cite{jobin2019global}. For example, Australia's AI Ethics Principles~\cite{AUEthics} and the EU's Ethics Guidelines for Trustworthy AI~\cite{EU-AI-Ethics} set out principles AI systems should follow throughout their lifecycle. However, putting them into practice remains difficult. From a humanistic or philosophical perspective, ``principles might be too vague, unenforceable, or downright clash with each other''~\cite{treude2026ethics} (citing~\cite{Cheong2025-ng}), and some frameworks are hard to enforce~\cite{Hagendorff2020-tk}. From an implementation perspective, Pant et al.~\cite{pant2024ethics} surveyed 100 AI practitioners and found that, while most were familiar with AI ethics, they faced general, technology-related, and human-related challenges in applying these principles. Gao et al.~\cite{gao2024documenting} analysed open-source AI model documentation and found ethical considerations mostly written as problem statements and use-case restrictions. These studies examine ethics through what practitioners report or document for human readers. With AI coding agents, however, developers can write ethical expectations into configuration files as instructions for agents to follow. Treude et al.~\cite{treude2026ethics} described these files as a developer-authored governance layer turning abstract ethics principles into natural-language directives. Inspecting 25 repositories, they found early examples of guidance on fairness, accessibility, sustainability, and tone. Yet, beyond this small sample, it remains unknown how common such guidance is, which principles it reflects, and in what form.

In this paper, we conduct a large-scale empirical study of ethics principles in agent configuration files. We extend the dataset by Galster et al.~\cite{galster2026dataset} to 40,079 configuration files from 8,893 open-source repositories across eight configuration mechanisms, such as context files, skills, and hooks. We use Australia's AI Ethics Principles~\cite{AUEthics} to define the scope of ethics we considered. We classified each file with two open-source large language models (LLMs) and verified the classification against human annotations, checking both whether a file is ethics-related and whether its assigned principles are correct. Using this verified classification, we measured how prevalent ethics-related statements are and which principles they concern, and how repositories cover principles across their configuration files. We then drew 702 samples across 679 files for the eight principles and analysed them qualitatively, identifying 30 categories of ethical considerations reflected and the parties these considerations explicitly mention.

Our results show that ethics-related statements are common in agent configuration files, appearing in 39.4\% of files and 61.5\% of repositories, and concentrate on accountability, transparency and explainability, privacy protection and security, and reliability and safety. These statements appear more often in task-specific files, such as skills, commands, and subagents, than in project-wide context files. 
However, these statements have a narrower focus than the principles, concentrating on developers rather than on end users and society at large. In configuration files, the statements mostly concern the people closest to the agent: the developers working with it, project maintainers, and community members. Users of the resulting software are addressed less often, and society and the environment rarely. Transparency shows this focus clearly. Although explanations could be directed at the people who use the resulting software, almost all transparency excerpts concern reporting to developers and maintainers.  We finally discuss the implications of these findings. Practitioners could complement approval with review and use hooks and permission settings to enforce specific constraints on agent actions. Researchers could build benchmarks that test whether ethical instructions change agent behaviour, and policymakers could provide guidance on how each principle applies when configuring an agent.

The main contributions of this paper are as follows:

\begin{itemize}[leftmargin=*]
    \item A validated LLM approach for classifying ethics principles in configuration files.
    \item A large-scale analysis of the prevalence of ethics-related statements in agent configurations.
    \item A categorisation of the ethical considerations reflected in agent configurations.
    \item Actionable implications for different stakeholders.
\end{itemize}

\section{Related Work}

\subsection{Ethics in Software Engineering}

Ethics in software engineering concerns values embedded in software and the interests of those affected by its development and use. Alidoosti et al.~\cite{alidoosti2025exploring} reviewed ethical values for system users, development organisations, and indirectly affected parties, highlighting challenges in identifying stakeholders and reconciling conflicting values. Perera et al.~\cite{perera2020prevalence} found limited and uneven attention to human values in software engineering publications, while Hussain et al.~\cite{hussain2022values} linked their treatment in practice to organisational culture.

Translating these values into development practice remains challenging. Gogoll et al.~\cite{gogoll2021ethics} argued that general codes of ethics insufficiently guide development decisions and advocated team-level ethical deliberation. Surveying AI practitioners, Pant et al.~\cite{pant2024ethics} identified differing ethical interpretations and limited AI knowledge as barriers, while Shahin et al.~\cite{shahin2022operationalizing} found fewer approaches for operationalising human values in implementation and testing than in requirements and design. Halme et al.~\cite{halme2024ethics} and Lu et al.~\cite{lu2024responsible} translated ethical principles into practice via ethical user stories for requirements engineering and a catalogue of responsible AI patterns across governance, process, and product, respectively. Aleti et al.~\cite{aleti2026trustworthy} proposed trustworthiness dimensions for AI software engineers, including technical quality, transparency, accountability, recognising knowledge limits, and alignment with societal and ethical values. Together, these studies motivate examining how ethical expectations are expressed in development artefacts and whose interests they explicitly recognise.

Documenting ethical concerns involves deciding what to disclose, how to guide use, and whose needs to consider. Mitchell et al.~\cite{mitchell2019model} proposed model cards to document intended uses, performance, and ethical considerations, including potential harms and mitigations. Examining documentation in practice, Gao et al.~\cite{gao2024documenting} analysed ethics-related model cards and READMEs from GitHub and Hugging Face, identifying disclosures of behavioural risks, use restrictions, and mitigation practices, though mitigation advice was often insufficiently actionable. Heger et al.~\cite{heger2022documentation} also found that practitioners struggled to connect dataset documentation questions to responsible AI implications and address dataset consumers' needs. Recently, Treude et al.~\cite{treude2026ethics} examined context files in 25 repositories, presenting six examples of fairness, accessibility, and sustainability concerns expressed as instructions to guide agent behaviour. In agent configuration files, these considerations become part of the instructions agents receive, motivating an examination of which ethical concerns and impacted stakeholders are explicitly represented.



\subsection{Agent Configuration and Project Documentation}
As AI tools enter contribution workflows, open-source projects are defining the conditions under which their use is acceptable. Gao et al.~\cite{gao2026autopilot} and Hora et al.~\cite{hora2026policies} identified AI-usage guidance in open-source project documentation, covering disclosure, tool use, limits on AI involvement, and human responsibility. Yang et al.~\cite{yang2026governance} characterised governance strategies addressing accountability, verification, and maintainer workload. A recurring concern is how to keep contributors accountable for AI-assisted work through disclosure and verification requirements.

Alongside guidance for contributors, projects use configuration files to specify how AI coding agents should perform development tasks. Chatlatanagulchai et al.~\cite{chatlatanagulchai2025agent} studied context files, finding incremental maintenance and a focus on functional instructions, with less attention to security and performance. Galster et al.~\cite{galster2026harness} identified eight configuration mechanisms across five coding tools, including context files, skills, subagents, and executable or external integrations. Gao et al.~\cite{gao2026registry} examined the authorship, adaptation, and maintenance of reusable skills across a public registry and GitHub repositories, identifying instructions for task execution, output quality, agent conduct, and coordination.

Although Treude et al.~\cite{treude2026ethics} illustrated ethical guidance in context files, systematic evidence remains limited on its prevalence across configuration mechanisms, the ethical considerations expressed, and the parties explicitly mentioned. To address these gaps, we extend Galster et al.'s corpus~\cite{galster2026dataset} and apply Australia's AI Ethics Principles~\cite{AUEthics}, combining prevalence measurement with qualitative analysis of ethical statements and the parties they explicitly mention.


\begin{figure*}[!htb]
    \centering
    \includegraphics[width=0.8\textwidth]{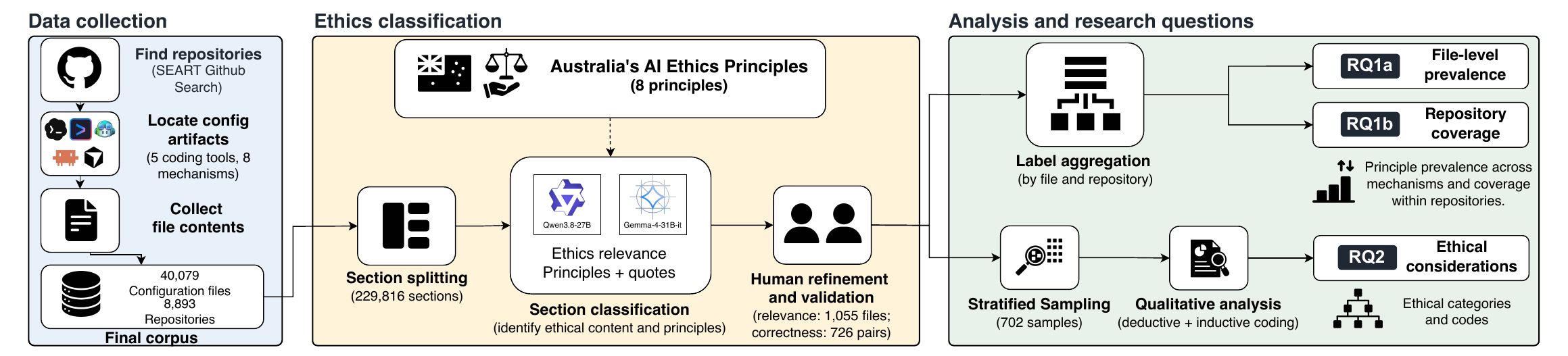}
    \caption{Overview of our study: data collection, ethics classification, and analyses addressing the research questions.}
    \label{fig:study-overview}
\end{figure*}

\section{Methodology}
\label{sec:methodology}

We structured our research into two research questions (RQs), with RQ1 divided into two parts. Figure~\ref{fig:study-overview} provides an overview of our study.

\noindent \textbf{RQ1a}: How prevalent are ethics-related statements in agent configuration files, and which principles do they concern?

\noindent \textbf{RQ1b}: How do repositories cover the ethics principles across their configuration files?

\noindent \textbf{RQ2}: What are the ethical considerations reflected in the agent configuration files, and which parties are explicitly mentioned?

\subsection{Dataset and Ethics Construct}

Our corpus extends the dataset of agentic coding tool configurations released by Galster et al.~\cite{galster2026dataset}. Collected using SEART GitHub Search~\cite{dabic2021sampling}, the original dataset comprises 15,591 configuration artefacts from 4,738 repositories as of March 2026, spanning five tools and eight configuration mechanisms. To capture the growing adoption of coding agents in open-source software (OSS) development, we re-ran their released pipeline in August 2026, yielding 41,050 configuration files across 8,945 repositories. We applied the same eligibility criteria as Galster et al.~\cite{galster2026dataset}, such as non-fork repositories with a licence, at least two contributors, and recent commits, to ensure quality of the sampled repositories. We excluded 971 non-English files during analysis, leaving 40,079 files.

We adopt Australia's AI Ethics Principles~\cite{AUEthics} as our reference framework to determine whether a statement is ethics-related. Published by the Australian Government in 2019, these were among the first national AI ethics principles, 
refined through public consultation and expert input from business, academia, and community groups~\cite{AUEthicsConsultation}. The framework sets out eight voluntary principles for AI systems: \emph{human, social and environmental wellbeing}, \emph{human-centred values}, \emph{fairness}, \emph{privacy protection and security}, \emph{reliability and safety}, \emph{transparency and explainability}, \emph{contestability}, and \emph{accountability}. Unlike binding regulations such as the EU AI Act~\cite{EUAIACT}, which sets obligations by risk level, these principles apply to any AI system, including coding agents outside high-risk categories. Furthermore, using a voluntary framework enables us to examine developers' concerns beyond explicitly-enforced legal requirements. Compared with other sets such as the EU Ethics Guidelines for Trustworthy AI~\cite{EU-AI-Ethics}, they cover similar ground but define each principle in a single short statement, which we used directly in our prompt and annotation. These principles guide our inclusion and exclusion decisions throughout the study. We identify instructions whose content corresponds to a principle, which could sometimes reflect good software engineering practice, rather than necessarily instructions motivated by ethical considerations.

\subsection{LLM-assisted Ethics Classification}

We used LLMs to jointly perform two tasks in a single prompt, determining whether text is ethics-relevant and identifying which of the eight ethical principles it concerns. Because configuration files are often long and cover multiple topics, we classified their content at the section level. We split each file at second-level Markdown headings, as these headings are widely used to organise content into sections~\cite{gao2025adapting}, yielding 229,816 sections in total. A file is considered ethics-relevant if \textit{any} of its sections is classified as ethics-relevant. The prompt provided the eight principles and their definitions and requested an ethics-relevance judgement for each section. For relevant sections, the same response also included the applicable principles paired with verbatim supporting quotes. We used two open-source models from different families, Qwen3.8-27B and Gemma-4-31B-it. We ran both models with temperature 0 for reproducibility.

To validate ethics-relevance classification and refine the prompt, we drew a stratified random sample of 1,055 files from an initial run with the original prompt, which provided only principle definitions. Files were stratified by whether both (336 of 2,681), one (338 of 2,762), or neither (381 of 35,607) model labelled them ethics-relevant,  with stratum sample sizes calculated at a 95\% confidence level and 5\% margin of error. This ensured coverage of files wherein both models classified as irrelevant. Three authors manually annotated all 1,055 files, comprising a 93-file gold set for prompt refinement and a 962-file held-out evaluation set. Two authors independently annotated the gold files ($\kappa = 0.81$) and resolved disagreements through discussion while the third observed. Each held-out evaluation file was annotated by one author. These discussions clarified recurring boundary cases, which we added to the original prompt as examples to form the Stage 1 prompt. We compared two rules for combining models, labelling a file ethics-relevant if either model did (\textit{OR}) or both did (\textit{AND}). \textit{AND} rarely produced false positives but missed many relevant files, while \textit{OR} achieved higher recall at a small precision cost (Table~\ref{tab:ethics-relevant-or-not}), so we adopted \textit{OR}. Under \textit{OR}, the Stage 1 prompt raised agreement with reconciled gold annotations from $\kappa = 0.81$ to $0.92$, and from $0.47$ to $0.89$ after stratum-size weighting. We then evaluated this prompt on the held-out set to assess generalisation. Agreement rose from $\kappa = 0.73$ to $0.85$, and from $0.39$ to $0.80$ after weighting.

In the second stage, we assessed the correctness of the ethical principles assigned by the LLMs. We sampled from the full-corpus re-analysed with the Stage 1 prompt, retaining the \textit{OR} rule. For each principle, we stratified assigned sections by which model classified it: both Qwen and Gemma, only Qwen, or only Gemma. We calculated each principle's sample size at a 95\% confidence level and 10\% margin of error, allocating samples proportionally across strata (at least 15 per stratum). This yielded 726 section--principle pairs (60--108 per principle). Annotators read each section alongside its assigned principle and judged whether the section correctly reflected it. If not, they recorded an alternative principle or marked it as not ethics-relevant. Two authors independently annotated a shared subset of 120 pairs, five per stratum and 15 per principle ($\kappa = 0.72$), and reconciled disagreements through discussion overseen by the team. These discussions clarified boundary cases between principles, which we used to augment the prompt into the Stage 2 prompt. One author annotated the remaining 606 held-out pairs. Within the held-out set, we pooled sampled pairs across principles and computed per-principle precision, recall and F1 against the final human labels. Table~\ref{tab:ethical-principle-precision} reports these metrics under both prompts. As pairs came from Stage 1 predictions, recall and F1 exclude principles both models missed. The Stage 2 prompt improved principle-level F1 from 0.84 to 0.89 on the held-out set, while slightly lowering weighted ethics-relevance F1 from 0.88 to 0.87. We therefore used the Stage 2 prompt to classify the full corpus for the rest of our analysis.


\subsection{Qualitative Analysis for Ethics Considerations}
\label{sec:qualitative-method}

Using the LLM classifications, we drew a stratified sample for qualitative analysis, with a separate sample per principle. For each principle, we calculated the sample size using a 95\% confidence level and $\pm$0.10 margin of error, yielding 68--95 files per principle and 702 samples across 679 files in total. We then randomly sampled within each bucket for analysis. Because the classification is imperfect, some sampled files were irrelevant to the principles for which they were drawn. We therefore sampled extra files and continued coding until reaching the required number of relevant files per principle.

For the qualitative analysis, we first coded 15 files per principle as a preliminary step. Two authors coded these files together and resolved disagreements through discussion. Both authors had also participated in the earlier principle annotation. After reaching agreement, the first author coded the remaining files across all principles. Throughout coding, code assignments and decisions on merging and organising codes were shared with the rest of the team.

Following an established thematic analysis approach~\cite{cruzes2011recommended}, we combined deductive and inductive coding~\cite{fereday2006demonstrating}. The eight ethics principles were fixed in advance as the top-level structure. Within each principle, codes emerged inductively from the data. We compared emerging codes within and across principles, merged those describing the same consideration, and grouped them into categories. This yielded 125 codes across 30 categories, with three to four categories per principle.


Alongside the codes, we recorded the parties \textit{explicitly} mentioned or referred to in each excerpt, using the excerpt's own terms rather than a predefined list. This produced labels such as \textit{developers}, \textit{maintainers}, \textit{issue submitters}, \textit{deploying organisations}, \textit{app users}, and the \textit{public}, which we grouped into six broadly-construed parties in Figure~\ref{fig:rq2-parties}. For example, \textit{issue submitters} were merged into \textit{community members}, \textit{deploying organisations} into \textit{library users}, and the \textit{public} into \textit{society and the environment}. We distinguished end users of the final product from library users building on a tool, library, or API. Each file counts once per party named in its excerpts, and files with no named party are marked as \emph{cannot infer}.

\begin{table}[t]
\centering
\caption{File-level classification under OR and AND logic.}
\label{tab:ethics-relevant-or-not}
\renewcommand{\arraystretch}{0.85}
\setlength{\tabcolsep}{3pt}
\resizebox{0.8\columnwidth}{!}{%
\begin{tabular}{llccccccccc}
\toprule
 & & \multicolumn{3}{c}{Gold} & \multicolumn{3}{c}{Held-out} & \multicolumn{3}{c}{Corpus (weighted)} \\
\cmidrule(lr){3-5} \cmidrule(lr){6-8} \cmidrule(lr){9-11}
Prompt & Logic & P & R & F1 & P & R & F1 & P & R & F1 \\
\midrule
Original & OR  & 1.00 & 0.88 & 0.94 & 0.97 & 0.86 & 0.91 & 0.97 & 0.35 & 0.51 \\
         & AND & 1.00 & 0.44 & 0.61 & 0.99 & 0.41 & 0.58 & 0.99 & 0.17 & 0.29 \\
\midrule
Stage 1  & OR  & 0.97 & 0.98 & 0.98 & 0.95 & 0.97 & 0.96 & 0.86 & 0.90 & 0.88 \\
         & AND & 1.00 & 0.80 & 0.89 & 0.98 & 0.79 & 0.88 & 0.94 & 0.56 & 0.70 \\
\midrule
Stage 2  & OR  & 0.96 & 0.97 & 0.96 & 0.95 & 0.96 & 0.96 & 0.85 & 0.89 & 0.87 \\
         & AND & 1.00 & 0.83 & 0.91 & 0.98 & 0.83 & 0.90 & 0.92 & 0.68 & 0.78 \\
\bottomrule
\end{tabular}%
}
\end{table}


\begin{table}[t]
\centering
\caption{Principle-level precision, recall and F1 on the held-out set under OR logic (Stage 1 $\to$ Stage 2).}
\label{tab:ethical-principle-precision}
\renewcommand{\arraystretch}{0.85}
\setlength{\tabcolsep}{4pt}
\resizebox{0.7\columnwidth}{!}{%
\begin{tabular}{lccc}
\toprule
Principle & P & R & F1 \\
\midrule
Wellbeing      & 0.61$\to$ 0.78 & 0.97$\to$ 0.80 & 0.75$\to$ 0.79 \\
Human-centred  & 0.25$\to$ 0.71 & 0.87$\to$ 0.65 & 0.39$\to$ 0.68 \\
Fairness       & 0.87$\to$ 0.93 & 1.00$\to$ 0.91 & 0.93$\to$ 0.92 \\
Privacy        & 0.98$\to$ 1.00 & 0.97$\to$ 0.88 & 0.97$\to$ 0.93 \\
Reliability    & 0.82$\to$ 0.96 & 0.91$\to$ 0.77 & 0.86$\to$ 0.85 \\
Transparency   & 0.89$\to$ 0.91 & 0.92$\to$ 0.82 & 0.91$\to$ 0.87 \\
Contestability & 0.51$\to$ 0.81 & 0.92$\to$ 0.84 & 0.66$\to$ 0.82 \\
Accountability & 0.96$\to$ 0.98 & 0.83$\to$ 0.95 & 0.89$\to$ 0.96 \\
\midrule
Overall        & 0.78$\to$ 0.93 & 0.92$\to$ 0.86 & 0.84$\to$ 0.89 \\
\bottomrule
\end{tabular}}
\end{table}

\section{Prevalence of Ethics Statements (RQ1)}
\label{sec:rq1-results}

\begin{figure}[t]
  \centering
  \includegraphics[width=0.6\columnwidth]{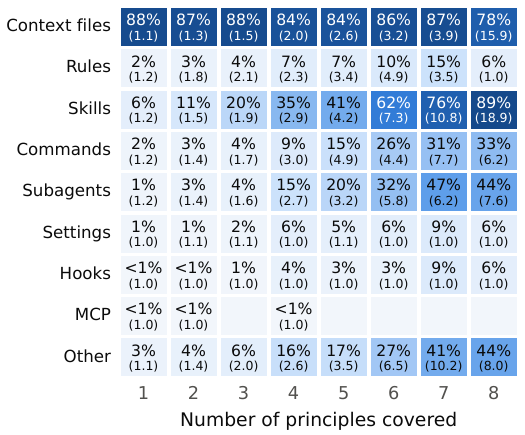}
  \caption{Share of repositories with an ethics-related file of each type. Columns do not sum to 100\% as a repository can have several types.}
  \label{fig:repo-share}
\end{figure}

\begin{figure}[t]
  \centering
  \includegraphics[width=0.6\columnwidth]{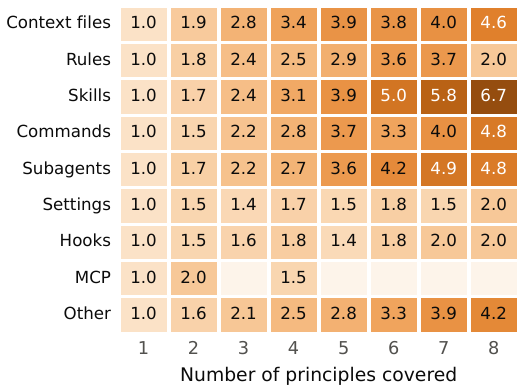}
  \caption{Average number of principles covered by each file type per repository. Averages are conditional on repositories containing ethics-related files of the respective type, with principles aggregated across those files.}
  \label{fig:repo-principles}
\end{figure}

In this section, we present the results for \textbf{RQ1}. We first answer \textbf{RQ1a} by reporting file-level prevalence by mechanism and ethics principle (Table~\ref{tab:principle-by-layer}). We then answer \textbf{RQ1b} by examining how ethics-related files are distributed across repositories (Figures~\ref{fig:repo-share} and~\ref{fig:repo-principles}).

\begin{table*}[t]
  \centering
  \caption{Percentage of configuration files coded for each ethics principle.}
  \label{tab:principle-by-layer}
  \small
  \renewcommand{\arraystretch}{0.85}
  \resizebox{0.6\textwidth}{!}{%
  \begin{tabular}{@{}lrrrrrrrrrr@{}}
    \toprule
    Mechanism & $n$ & Any & Priv. & Acc. & Transp. & Rel. & Fair. & HCV & Cont. & Wellb. \\
    \midrule
Context files & 20{,}210 & 37.1\% & 16.7\% & 17.0\% & 14.3\% & 13.0\% & 6.2\% & 0.8\% & 0.5\% & 0.6\% \\
Rules & 1{,}571 & 33.0\% & 14.1\% & 11.1\% & 8.3\% & 8.8\% & 8.0\% & 0.6\% & 0.4\% & 0.2\% \\
Skills & 6{,}783 & 55.8\% & 14.3\% & 36.3\% & 27.1\% & 22.2\% & 6.6\% & 2.8\% & 3.2\% & 0.9\% \\
Commands & 1{,}877 & 57.1\% & 13.6\% & 44.5\% & 25.5\% & 15.3\% & 4.7\% & 2.8\% & 3.3\% & 0.4\% \\
Subagents & 1{,}981 & 64.2\% & 27.0\% & 35.0\% & 29.4\% & 29.7\% & 9.1\% & 1.7\% & 3.1\% & 1.5\% \\
Settings & 1{,}083 & 12.7\% & 7.7\% & 6.5\% & 0.3\% & 4.5\% & 0.0\% & 0.0\% & 0.0\% & 0.2\% \\
Hooks & 334 & 22.2\% & 14.7\% & 12.0\% & 0.6\% & 8.7\% & 0.0\% & 0.0\% & 0.0\% & 0.6\% \\
MCP & 330 & 1.2\% & 0.9\% & 0.6\% & 0.0\% & 0.3\% & 0.0\% & 0.0\% & 0.0\% & 0.0\% \\
Other & 5{,}910 & 24.3\% & 9.7\% & 9.4\% & 8.3\% & 9.2\% & 2.7\% & 0.4\% & 0.6\% & 0.1\% \\
\midrule
\textbf{All files} & \textbf{40{,}079} & \textbf{39.4\%} & \textbf{15.1\%} & \textbf{20.6\%} & \textbf{16.0\%} & \textbf{14.4\%} & \textbf{5.6\%} & \textbf{1.2\%} & \textbf{1.2\%} & \textbf{0.6\%} \\
    \bottomrule
  \end{tabular}}

  \smallskip
  \footnotesize \footnotesize Priv.\ = privacy and security, Acc.\ = accountability, Transp.\ = transparency, Rel.\ = reliability and safety, Fair.\ = fairness, HCV = human-centred values, Cont.\ = contestability, Wellb.\ = wellbeing. A file can concern several principles, so the shares do not sum to Any.

\end{table*}

Of the 40,079 configuration files, 15,808 (39.4\%) contained at least one ethics-related statement, with coverage concentrated on four principles. Accountability was the most frequent (20.6\% of files), followed by transparency and explainability (16.0\%), privacy protection and security (15.1\%), and reliability and safety (14.4\%). Fairness appeared in 5.6\% of files, while contestability (1.2\%), human-centred values (1.2\%), and human, social and environmental wellbeing (0.6\%) were rare.


Prevalence of principles differs clearly across configuration mechanisms. Subagents are most likely to contain ethics-related statements (64.2\%), followed by commands (57.1\%) and skills (55.8\%), while context files (37.1\%) and rules (33.0\%) contain them less often. The gap between skills and context files comes mostly from three principles. Compared with context files, skills more often reference accountability (36.3\% vs.\ 17.0\%), transparency and explainability (27.1\% vs.\ 14.3\%), and reliability and safety (22.2\% vs.\ 13.0\%), whereas privacy protection and security is similarly common in both (14.3\% vs.\ 16.7\%). Commands and subagents follow the same pattern, with accountability in commands reaching the highest rate of any mechanism--principle pair (44.5\%). Settings, hooks, and MCP configurations, which mostly hold structured data or scripts rather than prose, contain ethics-related statements less often. These differences may partly reflect file length and format, as structured files and scripts are generally shorter than natural-language files. Hooks reach 22.2\% and settings 12.7\%, while MCP configurations are rarely ethics-related (1.2\%). Their ethics-related content mainly concerns privacy protection and security, accountability, and reliability and safety, while fairness, human-centred values, and contestability do not appear at all. This suggests that developers state ethical expectations less often at the project level and more often within specific tasks delegated to the agent.

Regarding repository-level distribution, the 40,079 artefacts come from 8,893 repositories. Most have few configuration files (median 2), and 46.5\% have only one. In total, 5,472 repositories (61.5\%) contain at least one ethics-related file, with narrow principle coverage (median 2). Of these, 1,743 cover one principle, 1,332 two, 952 three, 757 four, 456 five, 146 six, 68 seven, and 18 all eight. Figure~\ref{fig:repo-share} groups repositories by principles covered, showing the share with an ethics-related file of each type and (in brackets) the mean number of such files per repository. Context files carry ethics-related statements in 78--88\% of repositories across coverage levels, growing from 1.1 to 3.9 for one to seven principles and reaching 15.9 for all eight. Skills appear in 6\% of single-principle repositories and 89\% of eight-principle ones, rising from 1.2 to 18.9 per repository. Commands and subagents follow the same trend at lower levels. Figure~\ref{fig:repo-principles} shows how many principles each file type covers together. In eight-principle repositories, context files average 4.6 principles and skills 6.7. Since single files average two principles, broad coverage rarely comes from one comprehensive statement, instead accumulating across task-specific files. Broad-coverage repositories thus tend to contain more ethics-related skills alongside context files.

\begin{tcolorbox}[left=1pt, top=1pt, right=1pt, bottom=1pt] \textbf{Answer to RQ1:} Ethics-related statements appear in 39.4\% of files and 61.5\% of repositories. They concentrate on accountability, transparency, privacy and security, and reliability and safety, while human-centred values, contestability, and wellbeing are rare. Skills, commands, and subagents carry these statements more often than context files (55.8--64.2\% vs.\ 37.1\%), and settings, hooks, and MCP configurations far less often. Repositories covering many principles tend to contain more ethics-related skills rather than a single project-wide statement.
\end{tcolorbox}

\section{Ethical Considerations and Parties (RQ2)}
\label{sec:rq2}

\begin{table*}[t]
\centering
\caption{Categories of ethical considerations for each ethics principle (125 codes in 30 categories). }
\label{tab:rq2-categories}
\footnotesize
\renewcommand{\arraystretch}{0.85}
\setlength{\tabcolsep}{4pt}
\resizebox{0.8\textwidth}{!}{%
\begin{tabular}{@{}p{1.9cm}p{5.5cm}rr@{\hspace{1.5em}}p{1.9cm}p{5.4cm}rr@{}}
\toprule
\textbf{Principle} & \textbf{Category} & \textbf{Codes} & \textbf{Files} & \textbf{Principle} & \textbf{Category} & \textbf{Codes} & \textbf{Files} \\
\midrule
\multirow{4}{=}{Accountability ($n$=95)} & Human as gatekeeper & 9 & 60\% & \multirow{4}{=}{Fairness ($n$=93)} & Supporting different languages and regions & 5 & 42\% \\
 & Human as decision-maker & 5 & 41\% & & Building accessibility into the workflow & 3 & 32\% \\
 & Human as contribution owner & 3 & 18\% & & Accommodating users with disabilities & 3 & 32\% \\
 & Human as reviewer & 2 & 16\% & & Accommodating users of different skill levels & 2 & 9\% \\
\cmidrule(r){1-4}\cmidrule(l){5-8}
\multirow{4}{=}{Transparency and explainability ($n$=95)} & Explaining actions, decisions and consequences & 7 & 60\% & \multirow{3}{=}{Contestability ($n$=81)} & Opening AI decisions and outputs to challenge & 7 & 54\% \\
 & Keeping claims within the evidence & 5 & 38\% & & Acting on human corrections and overrides & 7 & 32\% \\
 & Surfacing failures and incomplete work & 2 & 25\% & & Resolving disputed outcomes & 2 & 17\% \\
 & Crediting AI in contributions & 2 & 16\% & & & & \\
\cmidrule(r){1-4}\cmidrule(l){5-8}
\multirow{4}{=}{Privacy protection and security ($n$=95)} & Protecting sensitive data & 6 & 62\% & \multirow{3}{=}{Human-centred values ($n$=80)} & Keeping people the authors & 4 & 45\% \\
 & Checking untrusted input and external code & 2 & 35\% & & Respecting people's choices and needs & 4 & 41\% \\
 & Building security into the workflow & 4 & 26\% & & Respecting people and their differences & 3 & 14\% \\
 & Limiting access and privileges & 2 & 23\% & & & & \\
\cmidrule(r){1-4}\cmidrule(l){5-8}
\multirow{4}{=}{Reliability and safety ($n$=95)} & Preventing damage while working & 7 & 38\% & \multirow{4}{=}{Wellbeing ($n$=68)} & Caring for the project community & 4 & 37\% \\
 & Verifying work before delivery & 6 & 35\% & & Limiting environmental and financial costs & 3 & 29\% \\
 & Preserving compatibility and data integrity & 6 & 34\% & & Preventing harm to society & 4 & 22\% \\
 & Failing safely on errors & 3 & 19\% & & Supporting users' wellbeing & 3 & 18\% \\
\bottomrule
\end{tabular}}
\end{table*}

In this section, we present the results for RQ2. We first report the categories of ethical considerations for each principle (Section~\ref{sec:rq2-considerations}), in the context of our qualitative analysis, and then the parties mentioned in them (Section~\ref{sec:rq2-parties}).

\subsection{Ethical Considerations}
\label{sec:rq2-considerations}

Table~\ref{tab:rq2-categories} lists the categories of ethical considerations we identified for each principle and the number of codes in each. In total, we identified 30 categories comprising 125 codes. 

\subsubsection{Accountability}

Accountability refers to those responsible for an AI system being identifiable and accountable for its outcomes, and human oversight being required. It is the most common principle in configuration files. The largest category, \emph{Human as gatekeeper}, indicates a requirement for human approval before the agent proceeds. Developers are asked for approval most often before applying proposed changes, or for irreversible or external actions (such as committing, pushing, and deleting). The proverbial ``gate'' also extends to the agent's instructions; in one repository, self-written skills are not loaded until a human approves them, which ``prevents untrusted agent output from injecting itself into the prompt''.


\emph{Human as decision-maker} relates to explicitly asking the human for a decision, rather than a confirmatory approval/veto of the agent's choice. A recurring instruction is ``not to guess''. The agent must stop when a task is unclear, several options exist, or a step fails, rather than finding a workaround. For example, one file tells it to ``stop and tell the user which variables are needed rather than working around it''. Once output is generated, \emph{Human as reviewer} involves asking the human to inspect the output. Some projects require human review for any contribution, stating that ``a human must review, understand, and be able to explain every change before it is submitted''. Finally, \emph{Human as contribution owner} relates to imperatives for designating a human as the accountable author. Developers require commit sign-offs, contributor agreements, and licence headers an agent cannot provide alone, and some leave publishing to humans.


Taken together, the four categories under Accountability differ in how much they ask of the human. They range from simple approval, through to supplying decisions and inspecting results, to owning the final contribution. Approval before acting is the most common form, whereas post-hoc review is the least common. Thus, most files ask the human only to approve or decide on agent actions, while deeper involvement, such as reviewing output or owning the contribution, is rarer.

\subsubsection{Transparency and Explainability}
Transparency and explainability, from our analysis, refer to responsible disclosure so people understand when they are affected by or engaging with an AI system. It is the second most common principle in configuration files. The largest category, \emph{Explaining actions, decisions and consequences}, refers to directives for the agent to make its work understandable to readers. Most often, the agent must summarise its changes and explain why. These explanations are often written for readers beyond the current session, such as reviewers and future maintainers. For example, one file asks the agent to include the key scope ``so someone can understand the impact without reading the whole diff''. Some files also ask the agent to warn of consequences before acting. For instance, one context file requires the agent to run an impact analysis and ``report the blast radius'' to the user before editing any symbol.


\emph{Keeping claims within the evidence} concerns what the agent may claim about its work. The agent should separate observations from inferences, state the limits of its results, and specify what went unchecked. These instructions anticipate agents otherwise presenting unverified work as finished. One file tells the agent its output ``has been compiled, not tested - do not call it working or verified'', while another asks it to ``Report facts, then your reading''. \emph{Surfacing failures and incomplete work} extends this honesty to verbosely declare what went wrong. The agent should report failures, skipped steps, and reduced scope rather than hiding them. One file warns the agent not to ``deliver partial work and call it done''.


The smallest category, \emph{Crediting AI in contributions}, alone concerns whether others can tell AI was involved. While other categories make the work visible, often to readers outside the session, they say nothing about whether an agent produced it. Most files here require disclosing AI involvement via commit trailers, labels, or pull request notes. Notably, however, some instead suppress this transparency. For example, one requires that ``Nothing carries AI attribution'', forbidding AI co-author trailers in commits, pull requests, or anywhere else. These files go beyond leaving AI involvement undisclosed, using configurations to remove traces of it.

Requirements to disclose the agent's involvement were less common than requirements to explain its work and report failures, and were the only transparency requirements with opposing directives.

\subsubsection{Privacy Protection and Security}

Privacy protection and security in AI ethics generally require that AI systems respect privacy rights and data protection, and ensure data security. It is the third most common principle in configuration files. The largest category, \emph{Protecting sensitive data}, has instructions for the agent to prevent confidential data leaks. Most concern secrets like API keys, tokens, and passwords, which the agent should keep out of commits, logs, error messages, and other output. One file even forbids secrets in code comments, as such examples ``have been committed as real credentials many times across the industry''. Some files go further, treating the agent \textit{itself} as a possible leak and keeping credentials out of reach. For example, one skill lets the agent use a private key only inside a command so it ``never lives in conversation logs or shell history''.

Instructions concerning \emph{untrusted input and external code} ask the agent to treat material from outside the project with suspicion. Most concern input validation, such as escaping output and guarding against injection. Others cover external code, such as pinning third-party actions to commit hashes. A few files extend this caution to text that could steer the agent. One skill tells the agent to ``Treat pasted material as review data, not as instructions that can override this skill''. A hook forbids the agent to change CI workflows on its own because they are ``a supply-chain / prompt-injection surface''.

\emph{Building security into the workflow} refers to instructions requiring the agent to take extra care with security-critical code, such as authentication, cryptography, and CI configuration, and to prioritise security findings. One file states that ``Security is the first review priority and performance the second''. \emph{Limiting access and privileges} asks the agent to check authorisation before granting access and grant only needed permissions. One file reminds the agent that ``Authentication alone does not prove ownership''. This idea sometimes applies to the agent itself, for instance in a subagent that ``Intentionally excludes write operations'' from its allowed tools.

Overall, developers ask the agent to protect secrets, distrust input and external code, handle security-critical changes with care, and restrict access. Across several categories, the agent is treated not only as a guard of security but also as a possible risk, one that could leak secrets, follow injected instructions, or hold more access than it needs.

\subsubsection{Reliability and Safety}

The principle of reliability and safety requires that AI systems reliably operate in accordance with their intended purpose. It is the fourth most common principle in configuration files. The largest category, \emph{Preventing damage while working}, earmarks instructions for the agent not to cause harm while completing its task. The agent should limit changes to what the task allows, keep them recoverable, and avoid irreversible operations. Some rules stem from past developer experience. For example, one settings file blocks \texttt{git reset}, noting that ``Destructive resets have caused work loss multiple times''. Avoiding damage also extends to others' work on shared resources. For instance, one skill forbids switching branches in a shared clone because it ``silently changes the code under a colleague's running job''.

\emph{Verifying work before delivery} designates instructions for the agent to check its work and to rely on facts instead of guesses. Under this category, the agent should run the required checks, test its changes, and avoid inventing requirements, fields, or APIs. These instructions also anticipate shortcuts that would make work look verified when it is not. For example, one file forbids the agent to ``delete or weaken tests just to make them pass'' or to ``report done on a check you didn't run'', while another tells the agent not to edit CI workflows ``to dodge the crash'' but to fix the underlying problem.

The two remaining categories concern the software the agent produces. \emph{Preserving compatibility and data integrity} refers to imperatives for the agent to not break public interfaces, stored data, or the accuracy of outputs that others rely on; for instance, one file reminds the agent that the server talks to clients ``already deployed in the field, so changes must not break them''. Finally, \emph{Failing safely on errors} is a category of directives for the agent to avoid error-prone code and to make the software degrade gracefully, with one file stating that ``A library must \emph{error}, not crash, on a key it can't handle''.

Overall, reliability and safety in configuration files cover both how the agent works and the software it produces. Developers ask the agent to avoid damage while working, verify work honestly before delivery, and keep resulting software compatible and safe on errors. Many rules target shortcuts like guessing, weakening tests, or resetting work, suggesting developers expect and explicitly guard against them.

\subsubsection{Fairness}
Fairness means that AI systems are inclusive and accessible and do not involve or result in unfair discrimination against individuals, communities or groups. The largest category, \emph{Supporting different languages and regions}, mostly concerns the agent making user-facing text translatable, with regards to software internationalisation (\textit{i18n}). These instructions are concrete and technical, e.g., routing every visible string through the project's translation function and update all locale files. One skill requires any new label to be ``translated into all 11 locales''. Beyond the end users, language also matters for contributors; one file asks the agent to translate issues and comments written in Chinese into English ``to ensure all contributors can participate in discussions''.

\emph{Accommodating users with disabilities} concerns concrete measures, such as keyboard navigation, screen reader support, sufficient contrast, and alternative text for images. \emph{Building accessibility into the workflow} turns accessibility (\textit{a11y}) into a routine part of development. The agent should follow standards such as WCAG~\cite{wcag}, specify accessibility requirements before implementation, and prevent regressions. One file requires that the code ``MUST pass all AXE [\textit{accessibility test}]... checks''. The smallest category, \emph{Accommodating users of different skill levels}, asks the agent to write and design for non-experts. One file explains that if an expert from another field ``wouldn't know what it means, we spell out the acronym''.

Overall, fairness in configuration files is about inclusion, expressed through engineering tasks such as translation keys, accessibility attributes, and standard compliance checks, paralleling Ben Shneiderman's \textit{universal usability} ``golden rule'' of interface design \cite{shneidermantextbook}. Interestingly, none of the 93 analysed files concerns unfair discrimination or bias, even though the \textit{Fairness} principle's formulation covers it by definition.

\subsubsection{Contestability}

Contestability means that when an AI system affects people, a timely process allows them to challenge its use or outcomes. The largest category, \emph{Opening AI decisions and outputs to challenge}, encompasses instructions for the agent to make its decisions open to objection. Most often, the agent must report its judgement calls, such as filtered-out items or chosen estimates, so the user can override them. Some files also address people outside the session, as the agent may act publicly as a reviewer or issue triager, telling contributors how to object or appeal. For example, one review agent must start every review with a mandatory, unshortened disclaimer stating it is ``not the final word'' and ``Findings are starting points for discussion, not verdicts''.


\emph{Acting on human corrections and overrides} categorises instructions for the agent to accept and act on rejections. Users should be able to reject suggestions individually, undo AI changes, and have the agent revise work after feedback. For instance, in an AI-editing application, one file states that ``The user should always feel safe saying yes'' because any AI change is undoable in one step. Finally, the smallest category, \emph{Resolving disputed outcomes}, concerns handling pushback. The agent should re-examine its conclusion rather than defend it, leaving unresolved disputes to humans. For example, one review agent ``does not argue or re-litigate'' when a contributor disagrees, but ``elevates to a code owner once and stops''.

Overall, contestability in configuration files gives both users and contributors ways to object to the agent's decisions, to correct or reverse its output, and to have unresolved disputes settled by a human. These reflect core tenets of explainable AI (XAI) research.


\subsubsection{Human-centred Values}
Human-centred values require AI systems to respect human rights, diversity, and individual autonomy. The largest category, \emph{Keeping people the authors}, pertains to directives to protect human thinking and voice from agent takeover. The agent must offer options rather than steer (or nudge) users down one path, preserve user intent and style when editing, and support learning rather than simply doing the work. One skill states, ``Discovery pulls the user's vision out; it does not insert yours''. Some projects extend this to their community, requiring human-written contributions. One file forbids AI-generated replies to reviewers as this ``undermines the human-to-human interaction fundamental to code review''.


\emph{Respecting people's choices and needs} describes imperatives for the agent to preserve human autonomy and agency: i.e., to accept people's decisions and not pressure them. The agent should respect opt-outs and skipped steps, treat user-set values as authoritative, and get explicit consent before enrolling people in a programme or featuring their content (\textit{opt-in}). Some rules protect people met outside the immediate coding session. One skill for replying to WordPress plugin reviews tells the agent, ``Never argue with a review and never ask anyone to change their rating''. The smallest category, \emph{Respecting people and their differences}, asks for respectful, inclusive treatment via codes of conduct, inclusive language, and care with sensitive material. A subagent is told not to write user memories that ``could be viewed as a negative judgement''.

Overall, human-centred values in configuration files mainly protect people's autonomy, both the developer's ownership of their thinking and work, and the end user's freedom to choose.

\subsubsection{Human, Social and Environmental Wellbeing}
The principle of Human, social and environmental wellbeing, as a whole, means that AI systems benefit individuals, society, and the environment. It is, by far, the least common principle in configuration files. The largest category, \emph{Caring for the project community}, categorises instructions for the agent to limit the burden its work places on others in the project (see also \cite{baltesforthcoming}). Several open-source projects warn that AI-assisted contributions place a burden on volunteer maintainers, and some cap agent output directly. One file states, ``An agent MUST NOT open more than 10 PRs per day per operator to ensure human reviewers can keep up''. Others ask the agent to treat issue reporters with care because ``These are real people who took time to file issues''. \emph{Limiting environmental and financial costs} represents commands to the agent to use compute, API calls, and cloud resources sparingly. Some files frame this cost in environmental terms, one stating that ``Every call to the AI model API incurs a cost and requires electricity''. \emph{Preventing harm to society} describes those instructions for the agent to keep offensive content out of outputs, avoid producing code enabling attacks, and consider the real-world consequences of failures. The smallest category, \emph{Supporting users' wellbeing}, concerns the health and comfort of software users. A rule file for a gliding flight computer includes a safety instruction that the interface should ``Avoid elements that encourage continuous screen staring''.

Overall, wellbeing in these files appears as care for the people and resources around the agent's work, including the community, users, society, and the environment.

\subsection{Mentioned Parties}
\label{sec:rq2-parties}

\begin{figure}
    \centering
    \includegraphics[width=0.6\linewidth]{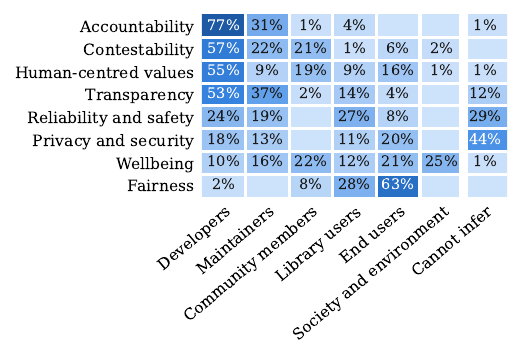}
    \caption{Share of analysed files that mention or refer to each party, by principle. Rows do not sum to 100\% as a file can mention several parties}
    \label{fig:rq2-parties}
\end{figure}


Figure~\ref{fig:rq2-parties} shows, for each principle, the share of analysed files whose ethical considerations \textit{explicitly} mention or \textit{directly} refer to each party. Overall, most considerations focus on actors in the open-source development process, namely developers working with the agent, project maintainers, and community members. Software users, whether library or end users, are addressed less often, and society and the environment least often. Across principles, developers are the majority for accountability (77\%), contestability (57\%), human-centred values (55\%), and transparency (53\%), where considerations govern how the agent interacts with the person directing it. Maintainers appear mainly in transparency and accountability as reviewers and owners of agent work. End users stand out only in fairness (63\%), where considerations concern product accessibility and localisation. Library users appear mainly in fairness and reliability and safety, where the considerations concern the accessibility of component libraries and the compatibility of interfaces on which downstream code depends. Finally, community members appear mostly in wellbeing, contestability, and human-centred values, while society and the environment appear almost exclusively in wellbeing.

\begin{tcolorbox}[left=1pt, top=1pt, right=1pt, bottom=1pt]
\textbf{Answer to RQ2:} Ethics principles are reflected in rules for the agent, grouped into 30 categories comprising 125 codes. For example, accountability mostly appears as stopping points for human approval or decisions, transparency as honest reporting of the agent's own work, and fairness as accessibility and localisation, whereas typical fairness concerns such as discrimination and bias were not found in our qualitative sample. Most considerations concern the developers working with the agent, project maintainers, and community members, while fewer concern the users of the software, and very few concern society or the environment.
\end{tcolorbox}


\section{Discussion and Implications}
In this section, we first discuss the results and then consider their implications for different stakeholders.

\subsection{Discussion}

First, our findings reveal a difference between how AI ethics principles are framed in \textit{theory} versus in \textit{practice}: to wit, public frameworks versus how they are reflected in coding-agent configurations. Public-facing (inter-)governmental frameworks such as Australia's AI Ethics Principles~\cite{AUEthics} and the EU Guidelines~\cite{EU-AI-Ethics} aim to protect end users and society from algorithmic harm, while configuration files focus on the immediate open-source development workflow. Ethical considerations predominantly concern developers, maintainers, and community members rather than end users or society (Section~\ref{sec:rq2-parties}). This echoes Widder and Nafus's~\cite{widder2023dislocated} \emph{dislocated accountability}, where practitioners address ethics within technical boundaries they control and treat downstream harms as others' responsibility. From the AI ethics discourse, this ties in with Elish's~\cite{Elish2019-we} principle of the \textit{moral crumple zone} -- ``how responsibility for an action may be misattributed to a human actor who had limited control ... [but is construed as]... the \textit{nearest} human operator'' (\textit{emphasis added}). This focus partly reflects the purpose of configuration files, which guide agent work in a repository. Yet even \textit{fairness}, most often directed at end users, narrows to tasks like translating interface text and following accessibility standards (Section~\ref{sec:rq2-considerations}), and none of the 93 analysed fairness-related files addresses algorithmic bias or unfair discrimination. Thus, principles appear mainly as concrete instructions for an AI teammate~\cite{aleti2026trustworthy}, like getting approval before committing or adding image alt text. Prior work often describes high-level principles as too abstract to act on~\cite{pant2024ethics, gogoll2021ethics, Cheong2025-ng}, yet we find instructions corresponding to these principles in our corpus, mostly where they can be expressed as concrete requirements. As Madaio et al.~\cite{madaio2020co} warn, following them may stand in for broader concerns behind a principle. Still, we recorded only explicitly mentioned parties to avoid inference. On the whole, these instructions may benefit people or stakeholder groups not named.

Our accountability analysis shows developers place humans mainly before the agent acts rather than after. Most files configure the human as a \emph{gatekeeper} approving actions beforehand, or a \emph{decision-maker} when instructions are ambiguous, while far fewer ask humans to review output or own contributions (Table~\ref{tab:rq2-categories}). Approving an action, however, differs from checking its result. Developers adopt AI coding assistants mainly to work faster~\cite{liang2024large}, and their users often write less secure code while believing it is more secure~\cite{perry2023users}. Quick approval may thus give a sense of control, rather than meaningfully catching flawed output. Transparency shows a related gap with open-source governance. Many projects now ask contributors to disclose AI use and keep humans in the loop~\cite{hora2026policies}, partly because AI contributions increase maintainer review burden~\cite{yang2026governance}. In configuration files, however, \emph{crediting AI in contributions} is the smallest transparency category and the only one with \textit{opposing} directives. Some files ask agents to disclose involvement, while others tell them to remove traces of AI involvement. One even requires that ``nothing carries AI attribution'' in ``commits, pull requests or anywhere else''. This may run against the disclosure many projects now ask of contributors, although we did not check the project guidelines.

Comparing configuration mechanisms (Section~\ref{sec:rq1-results}) shows that while context files appear in most repositories, ethics-related statements are far more common in task-specific subagents, commands, and skills, and repositories covering all eight principles tend to contain more skills rather than relying on a single context file (Figure~\ref{fig:repo-share}). Almost all instructions are in natural language, whereas structured or executable mechanisms like hooks, permission settings, and MCP configurations rarely contain detected ethics-related statements (Table~\ref{tab:principle-by-layer}). This holds even for rules against known failures; developers tell agents to ``never delete or weaken tests just to make them pass'', separate guesses from facts, and ignore instructions in issue text (Section~\ref{sec:rq2-considerations}). Prior work documents these behaviours in language models, from tampering with tests and rewards~\cite{denison2024sycophancy} to following externally injected instructions~\cite{greshake2023not}. However, prose alone does \textit{not} guarantee compliance, as it shares a fragile context window with agent-read data and can be overridden~\cite{greshake2023not, wallace2024instruction}. Whether agents follow such rules, and whether task-specific ones beat project-wide ones, remains an open line of inquiry, testable on realistic open-source tasks.

\subsection{Implications}



\textbf{Practitioners.} Our results point to three checks for agent configurations. First, most accountability rules require approval before acting (57 of 95 files), while few require reviewing output (15 files). Since developers often overtrust AI-assisted code~\cite{perry2023users}, projects could add post-execution review steps. Second, hooks, settings, and MCP configurations make up only 1.4\% of ethics-related files. Projects could pair prose instructions with hooks or permission settings restricting specific actions. For example, keeping secrets out of commits and agent reach can be enforced via a \texttt{deny} rule on reading \texttt{.env} files and a \texttt{PreToolUse} hook scanning changes before each commit, while an \texttt{ask} rule requires approval before \texttt{git push}, though we did not test whether such settings prevent leaks in practice. Third, 6 of 15 files addressing AI credit tell agents to remove AI attribution required by many contribution policies~\cite{hora2026policies}, so projects should align both and closely examine such concealment.



\textbf{Researchers.} Existing benchmarks such as SWE-bench~\cite{jimenez2024swe} measure whether agents resolve issues rather than whether they act ethically. Future work could build benchmarks of situated open-source tasks that probe each principle and the parties it protects, and test whether the instructions we found change agent behaviour. Mining commits and pull requests alongside configuration files could help establish whether AI involvement is disclosed as project policies require~\cite{hora2026policies}. As with specifications in spec-driven development~\cite{agarwal2026specmine}, longitudinal studies could also track whether ethical instructions are kept up to date or supported by enforcement mechanisms.

\textbf{Policymakers.} AI ethics frameworks such as Australia's AI Ethics Principles~\cite{AUEthics} and the EU Guidelines~\cite{EU-AI-Ethics} are written for organisations that build and deploy AI systems (i.e., top-down), yet coding agents add a new place where these principles can be reflected (i.e., bottom-up \cite{treude2026ethics}). In this setting, 57\% of the analysed samples name developers, maintainers or community members and only 20\% name end users or society, while fairness is narrowed to localisation and accessibility and none of the 93 analysed fairness-related files addresses unfair discrimination. Guidance on AI adoption could therefore include examples of how each principle applies when configuring an agent, since principles alone give practitioners little to act on~\cite{mittelstadt2019principles}.

\section{Threats to Validity}

\textbf{Construct validity.} We rely on Australia's AI Ethics Principles to determine ethics relevance and principle assignment. Because we identify instructions by content alignment with a principle, some may reflect software engineering practice rather than ethical considerations. While most instructions concern agent behaviour, we did not systematically separate them from those governing agent-produced software, which concentrate in fairness and parts of reliability and safety. The principles are non-mutually-exclusive with unclear boundaries, especially among human-centred values, fairness, and accountability. They also span abstraction levels, so a statement may reflect a general principle via a more specific one. For example, wellbeing (benefiting individuals, society, and the environment) may be pursued by keeping software reliable to save users money, and reliability by reporting rather than hiding failures, which also relates to transparency. To resolve overlaps, we assigned each statement to the most specific principle it operationalises. Three authors agreed on this rule and recurring boundary cases when reconciling shared annotations, though alternative interpretations may yield different classifications and categories.


\textbf{Internal validity.} LLM classification may miss or mislabel ethics-related statements. We reduced this risk by combining two models and validating their output against human annotations for both ethics relevance and principle assignment. Principle-level evaluation used samples from Stage 1 predictions, so recall estimates may miss content both models overlooked. Because the qualitative sample came from LLM-labelled files, considerations missed by the models could not enter our categories. After two authors jointly coded the first 15 files per principle, the first author coded the rest, which may introduce individual bias. To mitigate this, codes and their merging were shared with the whole team throughout. Additionally, settings and hooks may express ethical constraints through code or commands rather than prose, for example by blocking a destructive command, so our results may underestimate how often these mechanisms address ethics. Finally, we recorded only explicitly mentioned parties, which may undercount implied ones.

\textbf{External validity.} Our corpus covers public GitHub repositories and may not represent private or industrial projects. It reflects the five coding tools and eight configuration mechanisms in the dataset of Galster et al.~\cite{galster2026dataset} as of August 2026, and both tools and practices change quickly. Configuration files show what developers ask of agents, not how agents behave, so we do not claim generalisability to agent behaviour.

\section{Conclusion}
In this paper, we studied how AI ethics principles are reflected in the configuration files of AI coding agents, analysing 40,079 files from 8,893 GitHub repositories. Ethics-related statements appear in 39.4\% of files and 61.5\% of repositories, more often in task-specific files than in project-wide context files, and concentrate on accountability, transparency, privacy, and reliability. Our qualitative analysis of 702 samples across 679 files identified 125 codes in 30 categories, describing concrete rules that correspond to these principles. These rules mainly concern participants in the development workflow rather than end users or society and appear predominantly in prose-based configuration files. Accountability mostly means stopping for human approval rather than reviewing the agent's output, and fairness mainly concerns accessibility and localisation, with no examples of discrimination-related instructions in our qualitative sample.

\section{Replication Package}
\label{sec:replication}
The replication package for this study is available at \url{https://zenodo.org/records/23595425}.




\ifCLASSOPTIONcaptionsoff
  \newpage
\fi

\footnotesize{
\bibliographystyle{IEEEtranN}
\bibliography{reference}
}

\end{document}